\documentclass{llncs}

\usepackage[utf8]{inputenc}
\usepackage[T1]{fontenc}

\usepackage[english]{babel}

\usepackage[final]{pdfpages}

\usepackage{graphicx}
\usepackage{url}

\usepackage{multirow}
\usepackage{booktabs}

\usepackage[babel]{csquotes}

\usepackage{amsmath}
\usepackage{amsfonts}
\usepackage{amssymb}

\usepackage{listings}
\AtBeginDocument{}

\usepackage{listings}

\usepackage{hyperref}
\usepackage[capitalise,noabbrev]{cleveref}

\usepackage[misc]{ifsym} 

\newcommand{\ignore}[1]{} 

\def\mymath#1{\relax\ifmmode#1\else$#1$\fi}

\newcommand{\notsensible}{foolish}
\newcommand{\Notsensible}{Foolish}

\begin{document}

\title{Measuring Coverage of Prolog Programs Using Mutation Testing}
\author{Alexandros Efremidis\inst{1} \and Joshua Schmidt$^{1 \, \textrm{\Letter}}$ \and Sebastian Krings\inst{2} \and Philipp Körner\inst{1} \orcidID{0000-0001-7256-9560}}
\institute{
 Institut f\"{u}r Informatik, Universit\"{a}t D\"{u}sseldorf\\
  Universit\"{a}tsstr. 1, D-40225 D\"{u}sseldorf\\
\email{\{alefr101,joshua.schmidt,p.koerner\}@uni-duesseldorf.de}
\and
Niederrhein University of Applied Sciences \\
M\"onchengladbach, Germany\\
\email{sebastian.krings@hs-niederrhein.de}
 }

\maketitle
	
\begin{abstract}
  Testing is an important aspect in professional software development, both to
  avoid and identify bugs as well as to increase maintainability.
  However, increasing the number of tests beyond a reasonable amount hinders development progress.
  To decide on the completeness of a test suite, many approaches to assert test coverage have been suggested.
  Yet, frameworks for logic programs remain scarce.
  
  In this paper, we introduce a framework for Prolog programs measuring test coverage using mutations.
  We elaborate the main ideas of mutation testing 
  and transfer them to logic programs.
      To do so, we discuss the usefulness of different mutations in the context of Prolog
  and empirically evaluate them in a new mutation testing framework on different examples.
\end{abstract}

\section{Introduction \& Motivation}\label{introduction}
Testing is an important aspect in professional software development, both to
avoid and identify bugs as well as to increase maintainability.
However, tests themselves again consist of source code and possibly further artifacts that need to be maintained.
In modern software systems, code only needed for testing purposes can contribute between 33\%
and 50\% to the overall source code of a project~\cite{Sangwan:2006:TDL:1175884.1176039,Deursen01refactoringtest}.
In consequence, increasing the number of tests beyond a reasonable amount again hinders development progress.

The key to an efficient test suite is to assert the code coverage of existing tests.
That means, it is verified to what extent production code is tested.
Afterwards, one can remove tests that do not cover additional aspects and add tests where code is uncovered.
In \cref{sec:coverage-techniques},
we present different approaches to measure code coverage.

This paper makes two contributions:
Firstly, inspired by Toaldo and Vergilio~\cite{toaldo_applyingmutation},
we discuss several program transformations used for mutation testing
in \cref{section:mutation_operators}
and argue whether we deem them to be sensible or not.
Secondly, we implemented a framework featuring several mutations. Our framework is publicly available for SWI and SICStus Prolog and will be presented in \cref{mutation_implementation}.
In \cref{section:empirical}, this framework is used to evaluate
whether our intuitive classification of mutations
as sensible or \notsensible{} is correct 
by measuring the test coverage of several selected examples.
Finally, we discuss related and future work
in \cref{related,future}.

\section{Code Coverage Metrics}%
\label{sec:coverage-techniques}

Different metrics for code coverage, mostly differing in granularity, have been suggested and are at least partially applicable to Prolog:

\subsubsection{Predicate, Clause, Sub-Goal Coverage:}

A simple way to gain some insight into code coverage
is to execute tests
and trace which code was executed.
This can be done on different levels,
that is, on the level of sub-goals, clauses or predicates.
Moreover, there are different metrics to decide whether a specific level is covered.
For instance, a sub-goal, clause or predicate can be regarded as covered
if it succeeded at some point during execution.
In this paper, we use a more restricted metric.
We regard a clause to be covered if all sub-goals are covered, and, analogously, 
regard a predicate to be covered if all clauses are covered.
Success on each level can be traced, for instance,
by inserting tracing code via term expansion (source-to-source transformation) as done by Krings~\cite{kringsba},
using hooks into the Prolog interpreter (as done in SWI-Prolog's~\cite{DBLP:journals/corr/abs-1011-5332} testing framework PlUnit)
or executing the code in a meta-interpreter.

\subsubsection{Branch Coverage:}
Instead of considering individual program points covered if reached,
branch coverage considers if branching points such as if-statements 
have been executed both ways.
In particular, this implies that each condition has been evaluated to both true and false at least once.
For Prolog, this could be implemented either on the level of conditions,
but also on the level of individual calls.
In this case, we would expect the test suite to make each call fail and succeed at least once.
This would be harder to reach than the predicate coverage introduced above, given that each predicate would additionally have to fail at least once.
In case of Prolog, one also has to decide if and how redos of predicates should be counted.

\subsubsection{Path Coverage:}
Path coverage abstracts further from individual program points.
Instead of enforcing each condition to be evaluated in both directions,
path coverage considers all combinations of decision, that is, all paths through a predicate.
As above, one has to decide how redos should be counted, that is, if each combination of redo and later succeed is an individual path.

\subsubsection{MC/DC Analysis:}
A more sophisticated and popular approach
is coverage analysis via MC/DC (modified condition / decision coverage).
In order for code to be considered covered by MC/DC,
all conditions and decisions have to take all outcomes,
and, furthermore, each condition of a decision has to independently
influence the overall outcome of the decision.
MC/DC analysis can be implemented via term expansion as well.

\subsection{Mutation Testing}\label{mutation_testing}
The idea behind \textit{mutation testing} is to determine the test coverage by asserting the effectiveness of a test suite on modified versions of the source code under test.
To do so, syntactically nonequivalent versions of the source code, called \textit{mutants}, 
are generated which are intended to be semantically nonequivalent to the original code.
We view two programs to be semantically nonequivalent if they produce a different output for the same input at least once.
For instance, a mutant is generated by replacing an equality with an inequality.
Afterwards, all tests are executed.

If creating a semantically nonequivalent mutant, the semantics of the sources under test have changed.
We then expect at least one test to change its outcome.
That means, at least one positive test fails or negative test succeeds.
If this is the case, the mutant is called 
\textit{dead},
indicating that the test suite covers the mutated clause. 
Otherwise, the mutant is called 
\textit{alive},
indicating a lack of coverage.

To finally determine the coverage, a mutation testing tool will run the tests on the mutant, check the result and reset the mutant to the original code for the next iteration.
This cycle continues until no further mutation is possible.
Afterwards, the so-called \textit{mutation score} is computed.
That is, the number of dead mutants divided by the number of generated mutants.
The mutation score can be used as a measure for the test coverage.

As one can see, it is crucial to compute mutants that are indeed nonequivalent regarding their semantics,
as a semantically equivalent mutant will always be considered alive.
In consequence, the mutation score becomes less meaningful with an increasing amount of semantically equivalent mutants.
However, deciding whether two programs are semantically equivalent is, in general, undecidable~\cite{papadimitriou1985note}.

To counter this, 
it is possible to approximate the equivalence of two source code snippets 
by constraining the domains of used arguments, as for example suggested by Offutt and Pan~\cite{offutt1997automatically}.
Afterwards, their equivalence is checked exhaustively within these restricted domains.
Of course, this might result in detecting false positives
depending on the chosen domains.
For instance, two source code snippets might be semantically equivalent in a restricted domain
but have different behavior for values outside this specific domain.
Nevertheless, neglecting a mutation that results in a different semantics due to approximation
is superior to using a mutation that does not change the behavior at all.

In consequence, special care has to be taken when selecting mutation operators to be applied to the source code~\cite{grun2009impact}.
While it is only seldomly possible to find mutation operators without risk of generating semantically equivalent mutants,
the risk of semantically identical mutants differs between operators.

Libraries of useful mutations have been suggested for other languages,
such as the Javalanche framework~\cite{schuler2009javalanche}
and PIT~\cite{Coles:2016:PPM:2931037.2948707} for Java,
Mull~\cite{denisovmull} for LLVM,
and many more~\cite{jia2011analysis}.
However, due to the different nature of Prolog, those libraries cannot simply be adapted for mutation testing of Prolog programs.
In Prolog, for instance, negating an operator does not necessarily result
in semantically different code.
This can be seen in~\cref{fig:equiv-mutants}:
for instance, a predicate might treat some cases differently
in order to improve performance, 
like the empty list.
Mutating \verb|L = []| to \verb|L \= []|
will make the clause fail.
Yet, the Prolog interpreter will backtrack
and enter the second clause,
which may still find a solution for \verb|L = []|,
resulting in a semantically equivalent program.

\begin{figure}[t]
\begin{minipage}{0.49\textwidth}
\begin{lstlisting}
wrapped_sort(L, R) :-
    L = [], !,
    R = [].
wrapped_sort(L, R) :-
    sort(L, R).
\end{lstlisting}
\end{minipage}%
\begin{minipage}{0.49\textwidth}
\begin{lstlisting}
wrapped_sort(L, R) :-
    L \= [], !,
    R = [].
wrapped_sort(L, R) :-
    sort(L, R).
\end{lstlisting}
\end{minipage}
\caption{An Example for Equivalent Mutants via Negation}%
\label{fig:equiv-mutants}
\end{figure}

\section{Mutation Operators}
\label{section:mutation_operators}

In the following, we introduce our selection of mutation operators,
which is based on the selection suggested by Toaldo and Vergilio~\cite{toaldo_applyingmutation}.
We distinguish between \emph{sensible} operators,
which we expect to yield semantically different programs,
and \emph{\notsensible{}} operators,
were programs are expected to
retain the original semantics in most cases.
For most mutations,
examples that represent idiomatic Prolog code
are given.

In all cases,
we expect existing test cases to be reasonable:
for example, if the actual test initially fails,
backtracking should be avoided and the test should fail.
Furthermore, the test should compare with a (mostly) ground term
instead of allowing unification generously.
Test cases can either prove or disprove a goal.
When applying mutation testing, we expect all tests to succeed.
That means, a test disproving a goal should also succeed 
by checking for failure.

\subsection{Sensible Mutations}

From our experience, we consider the following transformations to be sensible:


\subsubsection{Predicate Removal:}
Deleting a predicate $\phi$,
more precisely all clauses of $\phi$ with the same arity,
is a sensible mutation because at least one test should fail, otherwise $\phi~$is not tested at all.
As long as $\phi~$is not dead code, the semantics change by removing $\phi$ due to the occurring existence error.
This mutation is comparable to predicate coverage,
as we expected at least one test to call $\phi$.

\subsubsection{Disjunction to Conjunction:}
By mutating a disjunction to a conjunction,
only a subset of queries can succeed:
now, they have to satisfy both branches.
Similar to branch coverage, we expect tests
to cover each branch individually.
In particular, there should exist a test
where the first alternative fails
whereas the second succeeds.
An example is given in \cref{fig:distocon}.

\begin{figure}[ht]
\begin{minipage}{0.49\textwidth}
\begin{lstlisting}
is_empty(L) :-
    L = [], !
    ;
    fail.
\end{lstlisting}
\end{minipage}
\begin{minipage}{0.49\textwidth}
\begin{lstlisting}
is_empty(L) :-
    L = [], !
    , 
    fail.
\end{lstlisting}
\end{minipage}
\caption{Changing Semantics by Replacing a Disjunction with a Conjunction.}
\label{fig:distocon}
\end{figure} 

In pure propositional logic,
replacing a disjunction with a conjunction
does not necessarily change the semantics
since a disjunction also provides the case where both arguments are true.
Prolog, on the other hand, does not execute the case that both disjuncts are true.
Instead, Prolog introduces a choice point leading to backtracking when searching for another solution.
This choice point is not retained by the mutant.
In practice, the calls within a disjunction often refer to the same variables providing alternative results.
To that effect, we expect replacing a disjunction with a conjunction
to alter the semantics in most cases, and, thus,
to be sensible in Prolog.

\subsubsection{Conjuntion to Disjunction:}
Replacing a conjunction by a disjunction does not necessarily change the semantics in pure propositional logic.
For instance, if $A \wedge B$ is true, the disjunction $A \vee B$ will be true, too.
However, if $A \wedge B$ is false, the disjunction will be true if $A$ or $B$ is true.
Therefore, depending on the interpretations of $A$ and $B$, 
there might be several cases in pure propositional logic 
where this mutation does not change the semantics.
In Prolog however, the data flow of a predicate often consists of passing arguments between predicates within a conjunction
as can be seen in~\cref{fig:conjunction_to_disjunction}.
In this context, replacing a conjunction by a disjunction will most likely change the semantics 
since the execution of this predicate will terminate after executing the first argument of a disjunction 
but initializing a choice point.
This also applies if a goal is supposed to fail.
\begin{figure}[ht]
\begin{minipage}{0.49\textwidth}
\begin{lstlisting}
flatten([L|Ls], FlatL) :-
    flatten(L, NewL),
    flatten(Ls, NewLs),
    append(NewL, NewLs, FlatL).
\end{lstlisting}
\end{minipage}
\begin{minipage}{0.49\textwidth}
\begin{lstlisting}
flatten([L|Ls], FlatL) :-
    flatten(L, NewL) ; 
    flatten(Ls, NewLs) ,
    append(NewL, NewLs, FlatL).
\end{lstlisting}
\end{minipage}
\caption{Semantics Change Caused by Replacing a Conjunction with a Disjunction.}
\label{fig:conjunction_to_disjunction}
\end{figure} 

\subsubsection{Atom or Variable to Anonymous Variable:}
Turning an atom or a variable to an anonymous variable causes
that certain values are no longer ground
which most likely changes the semantics.
Yet, tests may still pass if recursive cases are not tested.
Nevertheless, the predicate might be too complicated,
taking a variable as parameter that is not used within a clause.
Moreover, a variable might be a singleton one.
In both cases, replacing this variable by an anonymous one
will not necessarily change the semantics
of this specific clause as can be seen in~\cref{fig:problem_variable_to_anonymous}.
However, assuming that a clause does not define singleton or unnecessary variables,
this mutation will change the semantics in most cases.
These false positives still bear meaning about code quality,
where singleton variables 
and unnecessary parameters qualify as \enquote{code smell}
that should be avoided.

\begin{figure}[ht]
\begin{minipage}{0.49\textwidth}
\begin{lstlisting}
remove_dups([X,X|T],W,Res) :-
    remove_dups([X|T],W,Res).
\end{lstlisting}
\end{minipage}
\begin{minipage}{0.49\textwidth}
\begin{lstlisting}
remove_dups([X,X|T],_,Res) :-
    remove_dups([X|T],_,Res).
\end{lstlisting}
\end{minipage}
\caption{Predicate Retaining its Semantics when Replacing a Variable with an Anonymous One.}
\label{fig:problem_variable_to_anonymous}
\end{figure} 

\subsubsection{Interchanging Arithmetic Operators:}
When replacing two arithmetic operators with each other (e.g. replacing \verb|+| with \verb|*|),
a sensible mutant is likely to be created.
Since there are many operators to choose from
(e.g., an addition might be replaced by a subtraction, multiplication, division, modulo, \dots),
it is important to not create multiple mutants
to avoid an disproportional impact on the overall mutation score:
For instance, if an arithmetic operation is well tested, 
every mutation should fail.
Using several mutations would lead to a higher mutation score,
although the branch is already ensured to be covered by a single mutation.
On the other hand,
other branches without heavy arithmetic,
where only few sensible mutation are applicable,
would be valued lower due to the branch's lower amount of mutations.

\subsubsection{Interchanging Relational Operators:}
The mutation of relational operators is also sensible but not in every case.
It is not necessarily sensible to mutate \verb|A \== B| to \verb|A > B|
because multiple cases exist where the semantics do not change,
that is, every case where \verb|A > B| is true.
A sensible mutant is created by negating the relational operator (e.g. \verb|<| to \verb|>=|).
By negating a logical operator, the semantics is inverted:
every case which was true before will be false now and vice versa (see~\cref{fig:relop}).
\begin{figure}[ht]
\begin{minipage}{0.49\textwidth}
\begin{lstlisting}
min(A, B, A) :-
    A < B, !.
min(A, B, B).
\end{lstlisting}
\end{minipage}
\begin{minipage}{0.49\textwidth}
\begin{lstlisting}
min(A, B, A) :-
    A >= B, !.
min(A, B, B).
\end{lstlisting}
\end{minipage}
\caption{An Example for the Semantics Change of Relational Operators.}
\label{fig:relop}
\end{figure} 
\\

Negating a predicate likely manipulates the program's data flow due to the forced backtracking and possibly occurring side effects (see~\cref{fig:negpred}).
From a purely logical point of view, this mutation definitely changes the semantics.
\begin{figure}[ht]
\begin{minipage}{0.49\textwidth}
\begin{lstlisting}
rev([],[]).
rev([H|T],Rev) :- 
    rev(T,NT) , 
    append(NT,[H],Rev).
\end{lstlisting}
\end{minipage}
\begin{minipage}{0.49\textwidth}
\begin{lstlisting}
rev([],[]).
rev([H|T],Rev) :- 
    \+ rev(T,NT) , 
    append(NT,[H],Rev).
\end{lstlisting}
\end{minipage}
\caption{Example for Negating a Predicate Resulting in Different Semantics.}
\label{fig:negpred}
\end{figure} 

\subsection{\Notsensible{} Mutations}

There are also several mutations that we consider to be \notsensible{}.
Our reasoning is presented in the following:

\subsubsection{Clause Reversal:}
By reversing the order of clauses of a given predicate,
semantically different mutants are most likely just an infinite loop
in case a predicate is non-deterministic (see~\cref{fig:addlist}).
The creation of a non-terminating loop has no value for calculating the mutation score, 
because one cannot tell whether the mutated branch is tested or not due to the non-termination.
Therefore, we consider the result of a test that exceeds a reasonable time limit and a failing test to be different.
Otherwise, the mutation score would possibly be corrupted.

Furthermore, reversing the order of a predicate's clauses does not change the semantics in case the predicate is purely logical and deterministic, 
that is, each clause of the predicate validates its inputs (see~\cref{fig:revpred}).
\begin{figure}[ht]
\begin{minipage}{0.49\textwidth}
\begin{lstlisting}
add_to_list(L, _, 0, L).
add_to_list(L, E, C, R) :-
    CC is C - 1,
    LL = [E|L],
    add_to_list(LL, E, CC, R).
\end{lstlisting}
\end{minipage}
\begin{minipage}{0.49\textwidth}
\begin{lstlisting}
add_to_list(L, E, C, R) :-
    CC is C - 1,
    LL = [E|L],
    add_to_list(LL, E, CC, R).
add_to_list(L, _, 0, L).
\end{lstlisting}
\end{minipage}
\caption{An Example for a Non-Terminating Loop with Reversed Clause Ordering.}
\label{fig:addlist}
\end{figure} 

\begin{figure}[ht]
\begin{minipage}{0.49\textwidth}
\begin{lstlisting}
is_list([]).
is_list([_|T]) :-
    is_list(T).
 \end{lstlisting}
\end{minipage}
\begin{minipage}{0.49\textwidth}
\begin{lstlisting}
is_list([_|T]) :-
    is_list(T).
is_list([]).
\end{lstlisting}
\end{minipage}
\caption{An Example for an Equivalent Program with Reversed Clause Ordering.}
\label{fig:revpred}
\end{figure} 

\subsubsection{Cut Transformations:}
Inserting, removing and permuting cuts in Prolog is, in general, not a good idea. 
In Prolog, there are two kinds of cuts:
A cut is called \emph{red}, when its removal would create a semantically non-equivalent program.
Predicates with red cuts are, thus, not pure in a logical sense
and, per definition, behave differently without their cuts.
All other cuts are called \emph{green}:
while not affecting the semantics of the program,
they are used in order to increase performance.
In practice however, it is difficult to efficiently decide whether a cut is red or green.

In the context of mutation testing,
we only want to mutate red cuts in order to change the semantics.
Shifting a red cut to a subsequent position
is unlikely to change the semantics
since the condition of the cut has still been evaluated.
Thus, shifting red cuts to a prior position within the predicate
is most likely to be reasonable for mutation testing,
because the original condition after which a cut is called has not been evaluated.
By preventing backtracking, the semantics of the predicate is likely to be changed
as can be seen in~\cref{fig:cutprob}.

\begin{figure}[ht]
\begin{minipage}{0.49\textwidth}
\begin{lstlisting}
filter(_,[],[]).
filter(Pred,[H|T],[H|NT]) :- 
    call(Pred,H) , ! , 
    filter(Pred,T,NT).
filter(Pred,[_|T],NT) :- 
    filter(Pred,T,NT).
\end{lstlisting}
\end{minipage}
\begin{minipage}{0.49\textwidth}
\begin{lstlisting}
filter(_,[],[]).
filter(Pred,[H|T],[H|NT]) :- ! , 
    call(Pred,H) , 
    filter(Pred,T,NT).
filter(Pred,[_|T],NT) :- 
    filter(Pred,T,NT).
\end{lstlisting}
\end{minipage}
\caption{Changing the Semantics of a Predicate by Moving a Red Cut Forward.}
\label{fig:cutprob}
\end{figure} 

\section{A Mutation Testing Framework}\label{mutation_implementation}

In the following, we will take a closer look at the implementation of our framework.
As it relies on term expansion, the framework must be loaded before the module that shall be tested.
Before the tool is able to begin with the mutation testing process, some setup routines are executed:
In order to generate sensible mutants, the tool collects the source code's predicates upon term expansion.
Moreover, the term expander adds a \emph{dynamic} declaration for all predicates that may be modified.
The reason behind this is that the framework uses retraction and assertion to create mutants,
therefore the predicates have to be dynamic.
Then, the actual tests are executed on the original code
 to ensure that every test passes.
Otherwise, the criterion that tests pass on the mutant
is obviously flawed.
Furthermore, the overall runtime of the tests is stored 
in order to derive a reasonable timeout.
As mutations might result in infinite loops,
tests must be executed with a timeout on mutants.

\begin{figure}[th]
\centering
\includegraphics[scale = 0.48]{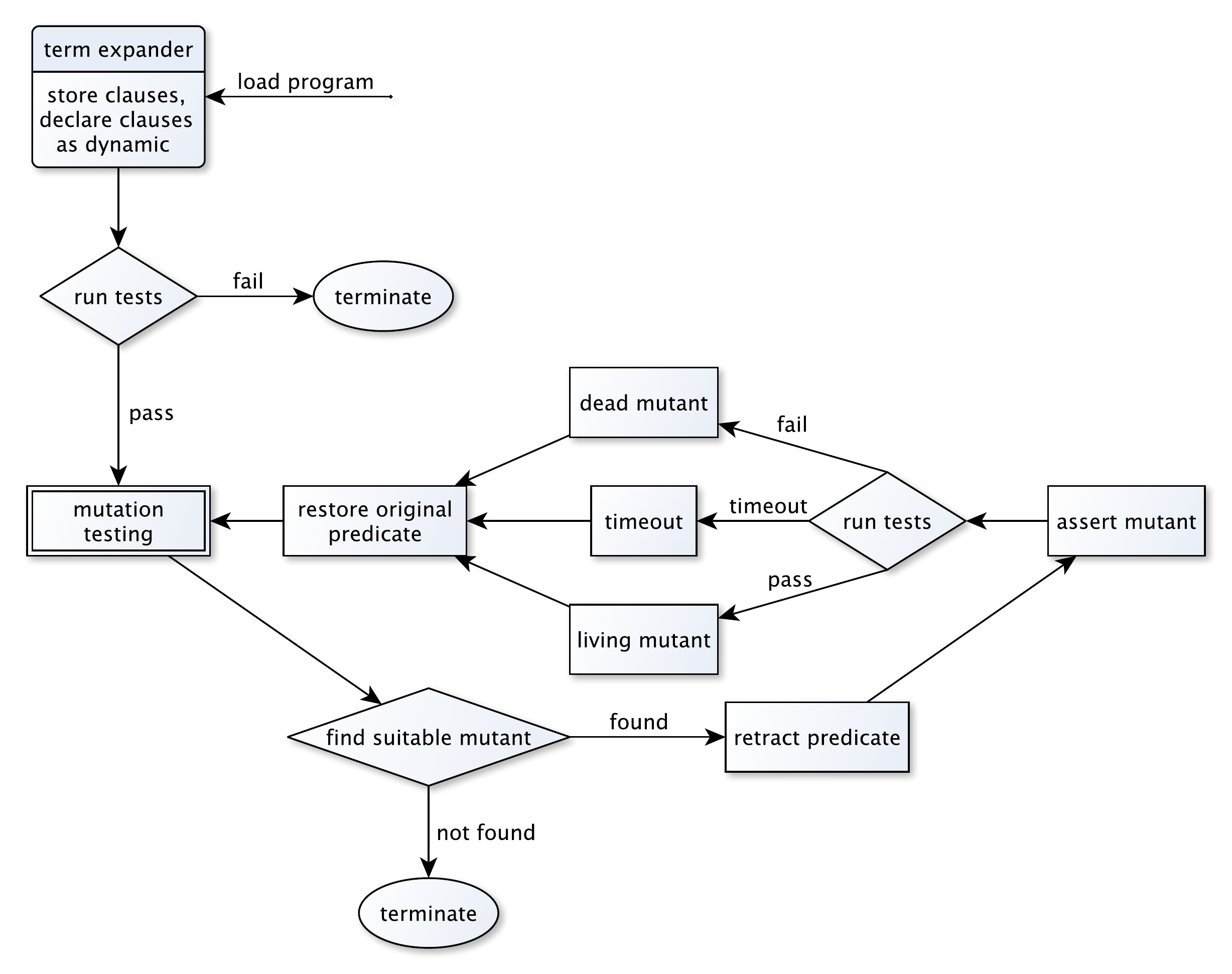}
\caption{The Framework's Workflow Diagram.}
\label{fig:flow}
\end{figure}

The process can be divided into the following procedures (also cf. \cref{fig:flow}): 
\begin{enumerate}
	\item find a suitable mutant, i.~e., a predicate where a new mutation is applicable
	\item generate a mutant
	\item retract the predicate and assert the mutant
	\item run the tests and check for failing tests or timeouts
	\item restore the original predicate
\end{enumerate}

This cycle continues until no other suitable mutation can be found.

Particular care has to be taken when manipulating a single clause of predicate:
the order of all clauses should (usually) be retained, but new facts can
only be asserted either at the top or bottom of the predicate.
In order to solve that problem,
the tool retracts and re-asserts the entire predicate, i.~e., all of its clauses, at once.

The mutation score is later calculated on the basis of all collected results.
Mutants are labeled either as \textit{dead}
when at least a single test has failed
or as \textit{alive} when all tests have passed.

However, a mutation may have caused an infinite loop.
To interrupt infinite loops,
tests on a mutant are executed with a timeout.
The timeout is defined as a constant plus the original runtime multiplied by two.
Mutants, which exceeded their test runtime are labeled as \textit{timeouts}.
In general, it is undecidable whether
an actual infinite loop was encountered
or the mutant just runs significantly longer.
To calculate the mutation score, mutants labeled as timeouts are not considered.
We thus deem the result of a test exceeding a time limit to be different from a failing test.
Otherwise, the mutation score might be corrupted 
in case of detecting false positives 
caused by a significantly longer runtime.
\\
Furthermore, this framework uses solely \textit{PlUnit}, in order to execute the tests, as an external library.
	
\section{Empirical Evaluation}
\label{section:empirical}

In this section, we aim to evaluate two different aspects of mutation testing.
Firstly, we verify our claims from \cref{section:mutation_operators}
by measuring living and dead mutants on several pieces of code
that we regard as reasonably tested.
Secondly, the overall mutation scores for these programs
are compared with coverage computed by predicate, clause and sub-goal coverage.
For the evaluation, we use several Prolog programs which can be found on GitHub\footnote{https://github.com/joshua27/mutation-testing-benchmarks}
along with a more detailed description.
Most of the programs are part of an evaluation
of different interpreter designs~\cite{kps15}.
We have chosen these programs 
since they are part of a publication 
and have been developed test driven.
Therefore, we expect these programs to have at least a mediocre test coverage.
Additionally, we test the coverage of a translation between
two formalisms (\verb|alloy2b|)~\cite{alloy2b}.
We think this program is interesting 
for a comparison of different coverage metrics 
since it only contains integration tests.
The code will thus not be tested in detail 
which probably has an impact on the different coverage scores.

\subsection{Sensibleness of Mutations}

A detailed overview of the results for our benchmarks
can be found in \cref{tbl:mutant-detailed},
where for each considered file,
the amount of living and dead mutants are given per mutation.

\begin{table}
    \setlength{\tabcolsep}{.5em}
    \caption{Overview of Living (First Number) / Dead (Second Number) Mutants in Real-World Examples. Timeouts are not considered.}%
    \label{tbl:mutant-detailed}

    \resizebox{\textwidth}{!}{
        \begin{tabular}{rrrrrrrr}
\multirow{2}{*}{Mutation} & \rotatebox[origin=c]{90}{ast\_interpreter} & \rotatebox[origin=c]{90}{compiler} & \rotatebox[origin=c]{90}{parser} & \rotatebox[origin=c]{90}{rational\_trees} & \rotatebox[origin=c]{90}{rt\_bytecode} & \rotatebox[origin=c]{90}{alloy2b} & \rotatebox[origin=c]{90}{eight\_puzzle}\\ \midrule
remove predicate & 0/3 & 4/11 & 0/5 & 0/5 & 0/4 & 8/68 & 4/17\\
\verb|;| to \verb|,| & 0/2 & 0/2 & 0/3 & 0/2 & 0/1 & 1/15 & 1/0\\
\verb|,| to \verb|;| & 0/30 & 10/34 & 1/14 & 4/12 & 12/31 & 38/245 & 22/22\\
\verb|=| to \verb|\=| & 0/1 & 0/2 & 0/0 & 0/1 & 0/0 & 4/51 & 1/0\\
\verb|\=| to \verb|=| & 0/0 & 0/0 & 0/0 & 0/0 & 0/0 & 0/1 & 0/0\\
\verb|=:=| to \verb|=\=| & 0/0 & 0/0 & 0/0 & 0/0 & 0/0 & 0/0 & 0/0\\
\verb|=\=| to \verb|=:=| & 0/0 & 0/0 & 0/0 & 0/0 & 0/0 & 0/0 & 0/0\\
\verb|==| to \verb|\==| & 0/2 & 0/0 & 0/0 & 0/3 & 0/1 & 1/2 & 0/0\\
\verb|\==| to \verb|==| & 0/0 & 0/0 & 0/0 & 0/0 & 0/0 & 0/0 & 0/0\\
\verb|>| to \verb|=<| & 0/0 & 0/0 & 0/0 & 0/0 & 0/0 & 0/2 & 1/2\\
\verb|>=| to \verb|<| & 0/0 & 0/4 & 0/7 & 0/0 & 0/0 & 0/0 & 0/0\\
\verb|<| to \verb|>=| & 0/0 & 0/2 & 0/0 & 0/0 & 0/0 & 0/1 & 0/1\\
\verb|=<| to \verb|>| & 0/0 & 0/0 & 2/5 & 0/0 & 0/0 & 0/0 & 0/1\\
\verb|+| to \verb|-| & 0/0 & 0/6 & 0/0 & 0/0 & 0/0 & 0/2 & 0/2\\
\verb|-| to \verb|+| & 0/0 & 4/35 & 0/0 & 0/5 & 0/0 & 0/1 & 0/2\\
\verb|*| to \verb|+| & 0/0 & 0/0 & 0/0 & 0/0 & 0/0 & 0/0 & 0/0\\
\verb|/| to \verb|-| & 0/0 & 0/0 & 0/0 & 0/0 & 0/0 & 0/0 & 0/0\\
increase number & 0/0 & 10/34 & 11/3 & 0/0 & 0/15 & 9/151 & 2/30\\
decrease number & 0/0 & 9/35 & 12/2 & 0/0 & 1/14 & 10/151 & 4/28\\
negate expression & 0/19 & 6/24 & 0/7 & 0/9 & 2/29 & 15/102 & 6/22\\
true to false & 0/3 & 0/2 & 0/0 & 0/2 & 0/0 & 1/2 & 0/0\\
false to true & 0/3 & 1/2 & 0/0 & 0/0 & 0/0 & 8/0 & 0/0\\
var to \verb|_| & 33/184 & 122/245 & 0/69 & 5/103 & 133/252 & 244/1327 & 95/138\\
atom to \verb|_| & 0/0 & 1/2 & 0/0 & 0/4 & 1/3 & 48/185 & 3/0\\
\verb|[]| to \verb|_| & 0/1 & 12/6 & 3/0 & 0/1 & 2/1 & 42/51 & 11/1\\
permute cut & 1/0 & 4/1 & 0/1 & 2/0 & 16/0 & 18/21 & 1/3\\
reverse predicate & 3/0 & 13/2 & 3/2 & 5/0 & 4/0 & 38/22 & 18/0\\
\end{tabular}

    }
\end{table}

As claimed,
removing a tested predicate always creates dead mutants.
In the cases where mutants are still alive,
there simply existed no test case for the predicates.
This is a mutation that is obviously sensible.

In our benchmarks, disjunctions have been fairly scarce;
this is due to the fact that usually
two separate clauses are preferred.
However, no mutants on the few disjunctions survived the testing,
so it seems to be as sensible as claimed.
Changing conjunctions to disjunctions, however,
generates more living mutants than expected.
This mutation is apparently not as sensible as assumed.

There are a few instances of living mutants
after negating a unification:
e.g., in the case of the \textit{eight\_puzzle\_solver},
the only unification unifies a potential solution to the actual solution
as a condition to terminate.
Since the implemented algorithms are expected to always find a solution,
terminating with a wrong solution after mutation is not caught.
Overall, it still seems to be a sensible mutation for most applications.

For the considered programs,
negating expressions results in a very small amount
of living mutants.
As this is fairly close to sub-goal coverage,
we think that they might be created
by uncovered code instead of false positives.

Changing variables to anonymous variables
results in a fairly large amount of mutants
in many cases,
yet the mutation is applicable
in a significantly larger amount of places as well.
There is a good chance that tests
do not cover all variables,
in particular where entire predicates
remain uncovered.
Overall, this mutation seems to be sensible as well.

As expected,
permuting cuts to an earlier position,
and reversing clauses of a predicate
often result is in semantically equivalent code.

Overall, our reasoning in \cref{section:mutation_operators} seems to be supported
by our measurements taken.
The only unexpected outcome is changing conjunctions to disjunctions.
Yet, for example, mutations concerning arithmetic
are not covered by our benchmarks
representatively.
For these transformations,
further code examples are required.

\subsection{Comparison with Predicate and Clause Coverage}

In the following, we will compare the coverage of 
different Prolog modules using the coverage tools of 
SWI\footnote{http://www.swi-prolog.org/pldoc/man?section=cover} and 
SICStus\footnote{https://sicstus.sics.se/sicstus/docs/4.3.2/html/sicstus/Coverage-Analysis.html} Prolog as well as the introduced mutation testing framework.
To give a short impression of the complexity of a program,
we list the number of predicates,
clauses and lines of code for each Prolog program in~\cref{table:evaluation}.

The coverage tool of SICStus Prolog measures 
how many times specific parts of the program, 
referred to as coverage sites, were executed.
According to the documentation of SICStus, 
a coverage site corresponds to all predicate calls 
like in \textit{trace} mode.
To that effect, there are different ways of interpreting the coverage results of the SICStus Prolog coverage analysis.
First, we compute the coverage on the level of clauses, that is, we view a predicate's clause to be uncovered if any coverage site within this clause is indicated to be uncovered.
Second, we compute the coverage on the level of predicates which we view as uncovered if they contain an uncovered clause.
Third, we compute a more detailed sub-goal coverage where we view each sub-goal independently.

The coverage tool of SWI Prolog behaves similar and computes the predicate coverage.
In the following, we will thus only refer to clause and predicate coverage without distinguishing between SWI and SICStus Prolog.

\begin{table}[]
\centering
\caption{Comparison of Prolog Coverage Tools}
\label{table:evaluation}
\begin{tabular}{l|r|r|r|r|r|r|r}
Prolog File & 
LoC & 
Predicates &
Clauses &
\parbox{1.5cm}{Clause Coverage} & 
\parbox{1.5cm}{Predicate Coverage} & 
\parbox{1.5cm}{Sub-Goal Coverage} & 
\parbox{1.5cm}{Mutation Coverage} \\ \hline
ast\_interpreter & 107 & 2 & 20 & 100.00\% & 100.00\% & 100.00\% & 88.10\%\\ \hline
compiler & 165 & 15 & 42 & 80.95\% & 62.50\% & 86.04\% & 52.60\% \\ \hline
parser & 80 & 38 & 16 & 100.00\% & 100.00\% & 100.00\% & 95.00\% \\ \hline
rational\_trees & 39 & 4 & 10 & 100.00\% & 100.00\% & 100.00\% & 93.70\% \\ \hline
rt\_bytecode & 105 & 4 & 33 & 93.93\% & 50.00\% & 95.24\% & 92.70\% \\ \hline
alloy2b & 725 & 74 & 198 & 84.41\% & 75.95\% & 87.68\% & 83.50\% \\ \hline
eight\_puzzle & 161 & 21 & 44 & 77.27\% & 67.10\% & 82.28\% & 78.50\%
\end{tabular}
\end{table}

For two source files, we encountered technical issues with our mutation testing framework.
Both programs rely on writing and consulting Prolog files at runtime.
Yet, when mutating the code,
the corresponding streams might not be closed properly,
resulting in an error caused by holding too many file handles simultaneously.

Overall, the results are non-binary.
In most cases, the results are similar
to the clause coverage approach.
For some files, our framework reports a higher score than
predicate coverage and, sometimes, even clause coverage.
Yet, it is able to find uncovered instances where both other approaches
claim perfect coverage.
In general, no approach can fully substitute all others.
Thus, we recommend to use mutation testing as an additional tool.

A rather unsatisfying result, however,
is that living mutants still require tedious, manual review
in order to find uncovered code and to verify that tests are missing.

\section{Related Work}\label{related}
As already stated in~\cref{section:mutation_operators}, different mutation operators for Prolog have been outlined by Toaldo and Vergilio~\cite{toaldo_applyingmutation}.
Our evaluation performed in~\cref{section:empirical} has shown that not all of these operators are efficient, for instance, they might generate numerous semantically identical mutations.

Of course, mutation testing can be performed on other, non-logical, languages as well.
Among the most prominent tools is PIT~\cite{Coles:2016:PPM:2931037.2948707}, a mutation testing tool for Java.
Imperative languages aside, mutation testing has been considered for declarative and functional languages as well~\cite{Alipour2014MutationTO}.
Usable tools exist, for instance, for Haskell~\cite{mucheck}. 

Regarding test coverage, several measurement tools are available and integrated into
the most common Prolog interpreters such as SWI~\cite{DBLP:journals/corr/abs-1011-5332} and SICStus~\cite{sicstusmanual}.
While they provide basic code coverage metrics, they usually only report on reached ports, that is, call, exit, redo and fail during execution.
As discussed in the introduction and shown in our empirical evaluation, this can lead to different results compared to calculating coverage based on mutation testing.
Of course, this does not imply that one metric is better than the other.

\section{Future Work}\label{future}
Even though we have improved the selection of mutation operators,
our approach still generates mutations semantically equivalent to the program under test.
As an equivalent mutation does not lead to a failing test by definition, each equivalent mutation causes a false-negative to be reported.
With our current implementation, deciding whether a mutation is semantically equivalent is done after the test results are reported.
In particular, the decision is made manually by the programmer.

To improve the efficiency of our test framework, techniques to detect semantically equivalent mutations should be incorporated in the future.
For instance, we are able to approximate the check for equivalence of two programs 
by restricting the domains of the arguments and 
using constraint solving to search for a counter example.
However, Prolog is a dynamic language not providing types which hampers constraint solving.
To counter this, we could try to detect the types of a predicate's arguments at runtime.
Unfortunately, this is not possible in general, 
for instance, if an argument is a variable which is not unified within a specific predicate call.
\textit{plspec}~\cite{plspec}, for example, offers a simple and easily extensible domain specific language for type annotations.
If a predicate is annotated using \textit{plspec}, we are able to derive the types of arguments as well as their role, that means, whether they act as an input or output of a predicate.
One downside of approximating the equivalence of two programs 
is that we do not consider possible side-effects of a predicate.
Nevertheless, in practice, 
we expect an approximation with appropriate domains 
will exclude more false negatives than true negatives.
To that effect, the meaningfulness of mutation testing will probably increase.

Another future work is to integrate automated test case generation in the mutation testing framework.
Again, we need to be able to derive the types of the arguments of a predicate and assume that \textit{plspec} has been used.
When running mutation testing, we gain a lot of information about the program under test.
In case a mutation does not cause a failing test, 
that means the mutant is alive, 
we might be able to use the mutated code together with the original code 
to generate an appropriate test case covering the mutated clause.
For instance, we could mutate a predicate $p$ using two arguments while the first argument acts as an input and the second one as an output.
The predicate is mutated to $p_{mut}$ and the mutation stays alive.
Assuming the mutation did change the semantics of $p$, 
we know that our test case needs to satisfy a call to $p$ and has to fail for $p_{mut}$.
We can then use techniques like fuzzing to generate randomized inputs until a set of parameters has been found.
Furthermore, we can use constraint solving to search for appropriate arguments.
Doing so, we can, on the one hand, search for any arguments that cause different behavior on the level of the predicate call.
That means, the arguments satisfy the constraint $\exists a,b: p(a,b) \wedge \neg p_{mut}(a,b)$.
On the other hand, in case we know which arguments are inputs and outputs, we can probably generate a more detailed test case by searching for input values that cause different output values.
In the context of our example that results to asserting $\exists a: p(a,b) \wedge p_{mut}(a,\bar b) \wedge b \neq \bar b$ to hold.
Of course, the implementation of the predicate $p$ might be faulty.
We thus can not automatically determine a generated test case to be correct.
However, we can present generated test cases to the user who can validate the behavior.

\section{Conclusion}\label{conclusion}
In conclusion, we have presented a framework for performing mutation testing on Prolog code.
Starting from the discussion of mutation rules by Toaldo and Vergilio~\cite{toaldo_applyingmutation},
we have devised a set of mutation rules we deem sensible,
i.~e., we suspect them to mostly compute semantically different mutations.
Our testing framework is available both for SICStus and SWI Prolog and can be downloaded from \texttt{https://github.com/hhu-stups/prolog-mutation-testing}.

We have shown empirically, that our mutation testing framework can handle different Prolog programs.
In particular, we tested it on larger examples, showing both applicability and performance of our approach.
Furthermore, we have shown that mutation testing indeed reports different coverage statistics than the ones provided by the coverage analysis tools shipping with SICStus and SWI.
We do not want to start a discussion on whether predicate coverage, path coverage or MC/DC analysis is better or worse than mutation testing.
Instead, we argue that any further knowledge about test coverage and the validity of a test suite helps to improve the overall implementation.

\bibliographystyle{abbrv}
\bibliography{paper}

\end{document}